%Paper: hep-th/9303060
%From: FHALON@WISWIC.WEIZMANN.AC.IL
%Date: Wed, 10 Mar 1993 10:07:30 GMT

%
% Printing instructions:
%       This paper needs the macro packages phyzzx.tex and tables.tex
%       The tables should be stripped off and printed separately.
%
%
\input phyzzx
\tolerance=10000
\sequentialequations
\def\rl{\rightline}

\def\t1{{\tilde 1}}

\def\AEF{A.E. Faraggi}
\def\DVN{D. V. Nanopoulos}

\def\NPB#1#2#3{Nucl. Phys. B    {\bf#1} (19#2) #3}
\def\PLB#1#2#3{Phys. Lett. B    {\bf#1} (19#2) #3}

\def\MODA#1#2#3{Mod. Phys. Lett. A {\bf#1} (19#2) #3}
\def\IJMP#1#2#3{Int. J. Mod. Phys. A {\bf#1} (19#2) #3}

\REF\GSW{M. Green, J. Schwarz and E. Witten,
Superstring Theory, 2 vols., Cambridge University Press, 1987.}
\REF\EU{\AEF, \PLB{278}{92}{131}.}
\REF\TOP{\AEF, \PLB{274}{92}{47}.}
\REF\SLM{\AEF, \NPB{387}{92}{239}.}
\REF\FFF{I. Antoniadis, C. Bachas, and C. Kounnas, Nucl. Phys. {\bf B289}
(1987) 87; I. Antoniadis and C. Bachas, Nucl. Phys. {\bf B298} (1988)
586; H. Kawai, D.C. Lewellen, and S.H.-H. Tye, Phys. Rev. Lett. {\bf57} (1986)
1832; Phys. Rev. {\bf D34} (1986) 3794; Nucl. Phys. {\bf B288} (1987) 1;
R. Bluhm, L. Dolan, and P. Goddard, Nucl. Phys. {\bf B309} (1988) 330.}
\REF\GCU{\AEF, WIS--92/17/FEB--PH, Phys. Lett. {\bf B}, in press.}
\REF\revisited{I. Antoniadis, J. Ellis,J. Hagelin, and \DVN,
\PLB{194}{87}{231}; G.K. Leontaris and J.D. Vergados, \PLB{258}{91}{111}.}
\REF\NM{\AEF, \PLB{245}{90}{437}.}
\REF\REVAMP{I. Antoniadis, J. Ellis, J. Hagelin, and \DVN, \PLB{231}{89}{65}.}
\REF\DSW{M. Dine, N. Seiberg and E. Witten, Nucl. Phys. {\bf B289} (1987) 585.}
\REF\ADS{J.J. Atick, L.J. Dixon and A. Sen, \NPB{292}{87}{109};
S. Cecotti, S. Ferrara and M. Villasante, \IJMP{2}{87}{1839}.}
\REF\KLN{S. Kalara, J. Lopez and D.V. Nanopoulos,
\PLB{245}{91}{421}; \NPB{353}{91}{650}.}
\REF\NRT{\AEF, WIS--92/48/JUN--PH, Nucl. Phys. {\bf B}, in press.}
\REF\GEN{\AEF, WIS--92/81/OCT--PH.}
\REF\CAB{A.E. Faraggi and E. Halyo, WIS 93/3/JAN-PH.}
\REF\FN{A.E. Faraggi and \DVN, \MODA{6}{92}{61}.}

\singlespace
\rl{WIS--93/4/JAN--PH}
\rl{SSCL--Preprint--193}
\rl{\today}
\rl{T}
\normalspace
\smallskip
\titlestyle{\bf{Neutrino Masses in Superstring
Derived  Standard--like Models}}
\author{Alon E. Faraggi{\footnote*{e--mail address: fhalon@weizmann.bitnet}}}
\smallskip
\centerline{SSC Laboratory, 2550 Beckleymeade Ave., Dallas TX 75237}
\centerline{and}
\centerline{Department of Physics, Weizmann Institute of Science}
\centerline{Rehovot 76100, Israel{\footnote\ddag{Permanent address}}}
\smallskip
\centerline{and}
\centerline{Edi Halyo {\footnote\dag{e--mail address:
jphalyo@weizmann.bitnet}}}
\smallskip
\centerline {Department of Physics, Weizmann Institute of Science}
\centerline {Rehovot 76100, Israel}
\titlestyle{ABSTRACT}
We propose a new scenario in a class of superstring derived standard--like
models that explains the suppression of the left--handed neutrino masses.
Due to nonrenormalizable terms and the breaking of the $U(1)_{Z^\prime}$
symmetry a generalized see--saw mechanism takes place. Contrary to the
traditional see--saw mechanism in GUTs, the see--saw scale and the
right--handed neutrino mass scale are suppressed relative to the
$U(1)_{Z^\prime}$ breaking scale.

\nopagenumbers
\pageno=0
%\singlespace
\endpage
\normalspace
\baselineskip=20pt
\pagenumbers
\centerline{\bf 1. Introduction}
Superstring theories  [\GSW] are believed to provide a viable framework for the
consistent
unification of gravity with the gauge interactions.
The superstring unification scale is at the Planck scale while at the
electro--weak scale
the Standard Model is in remarkable agreement with experimental
observations.
Thus, an extremely important task is to try to derive the Standard
Model from the superstring theories.

In ref. [\EU,\TOP,\SLM], realistic superstring standard--like models were
constructed in the free
fermionic formulation [\FFF] with the following characteristics:

\parindent=-15pt

1. Three and only three generations of chiral fermions.
There are no additional generations and mirror generations that
presumably get massive at a high scale.

2. The gauge group is $SU(3)\times {SU(2)}\times{U(1)_{B-L}}\times{U(1)_Y}
\times{U(1)^n}\times {hidden}$.

3. There are enough scalar doublets and singlets to break the symmetry in a
realistic way and to generate a realistic fermion mass
hierarchy.

4. Proton decay from dimension four operators is supressed due
to the gauged $B-L$
symmetry. There is a stringy doublet-triplet splitting mechanism that
projects out the color triplets and leaves the Higgs doublets
in the spectrum. Thus, proton decay from dimension five
operators is expected to be supressed [\SLM].

5. String gauge coupling unification can be obtained in these models due to
the presence of additional color triplets in vector--like
representations from exotic sectors [\GCU].

6. These models suggest an explanation for the top--bottom quark mass
hierarchy and for the generation mass hierarchy in general. At the
trilinear level of the superpotential only the top quark gets a
nonvanishing mass. The mass terms for the bottom quark and for the
lighter quarks
and leptons are obtained from nonrenormalizable terms.
Thus, only the top mass is characterized by the electroweak scale and
the masses for the lighter quarks
and leptons are naturally supressed [\TOP,\SLM].

\parindent=15pt

In this paper we examine the problem of neutrino masses
in the superstring standard--like models.
Like many other extensions of the Standard Model the superstring
standard--like models contain right-handed neutrinos.
At trilevel an underlying $SO(10)$
symmetry dictates that for every up--like quark mass term there is a
neutrino Dirac mass term,
with $\lambda_\nu={\lambda_u}$.
Thus, the $\tau$ neutrino mass is equal to the top quark mass.
Clearly a disasterous relation, unless there is a see--saw type
mechanism.
In this paper we demonstrate that a see--saw mechanism is
present in the superstring standard--like models.

We study the renormalizable and nonrenormalizable
contributions to the neutrino mass matrix.
We show that a large see--saw in the neutrino mass matrix
is generated by condensates
of a non--Abelian hidden gauge group.
As a result the left--handed neutrino
masses are supressed whereas the right--handed neutrino masses are pushed
above
the TeV scale. However, we find a
new feature in our superstring derived see--saw
mechanism unprecedented in traditional Grand Unified Theories (GUTs).
In traditional superstring inspired GUTs the see--saw scale is proportional
to the GUT scale. For example in the flipped $SU(5)$ the see--saw matrix
takes the form [\revisited],
$${\left(\matrix{{\nu_i}&{\nu_j^C}&{\phi_m}}\right)}
  {\left(\matrix{0&{(km_{_U})_{ij}}&0\cr
               {{(km_{_U})_{ij}}&0&M_G}\cr
               {0&M_G&O(M_W)}\cr}\right)}
               {\left(\matrix{{\nu_i}  \cr
                              {\nu_j^C}\cr
                              {\phi_m} \cr}\right)}, \eqno(1) $$
where $m_u$ is the up quark mass matrix, $M_G$ is the scale of
$SU(5)\times U(1)$ breaking. The mass eigenvalues are
$m_\nu\sim\langle\phi\rangle({{k m_u}\over{M_G}})^2$ and
$m_{N,\phi}\sim{M_G}$. Thus the mass scale of the right--handed neurtrinos
is the GUT scale. Similarly in the superstring inspired
standard--like model [\NM], the see--saw scale and
the mass of the right--handed neutrinos are proportional to the scale
of $U(1)_{Z^\prime}$ breaking.

In the superstring derived see--saw mechanism, the $U(1)_{Z^{\prime}}$ is
broken by
hidden sector condensates. We show that although the scale of the hidden sector
condensates is high, $\Lambda_h\sim10^{14}~GeV$, the masses of
the right--handed neutrinos are suppressed relative to the $U(1)_{Z^\prime}$
breaking scale.

\bigskip
\centerline{\bf 2. The superstring model}

To illustrate the superstring induced see--saw mechanism,
we examine the model of Ref. [\EU]. This model is constructed in the
free fermionic formulation [\FFF],
with a set of eight boundary condition vectors
for all the world--sheet fermions.
The first five vectors in the
basis consist of the NAHE{\footnote*{This set was first
constructed by Nanopoulos, Antoniadis, Hagelin and Ellis  (NAHE)
in the construction
of  the flipped $SU(5)$ [\REVAMP].  {\it nahe}=pretty, in
Hebrew.}} set, $\{{{\bf 1},S,b_1,b_2,b_3}\}$
[\REVAMP,\SLM].
In addition to the first five
vectors, the basis contains three additional vectors.
These vectors and the choice of generalized GSO projection coefficients
are given in Table 1, where the notation of Ref. [\SLM] is used.

The gauge group after application of the generalized GSO
projections is

 Observable{\footnote\dag{$U(1)_C={1\over2}U(1)_{B-L}$,
         $U(1)_L={1\over2}U(1)_{T_{3_R}}$.}}
: $SU(3)_C\times U(1)_C\times SU(2)_L\times U(1)_L\times U(1)^6$

 Hidden{\footnote\#{Hidden here means that the states which are
identified with the chiral generations do not transform under the hidden
gauge group.}}
\hskip 0.4cm    :   $SU(5)_H\times SU(3)_H\times U(1)^2.$

The weak hypercharge is uniquely given by
$U(1)_Y={1\over3}U(1)_C+{1\over2}U(1)_L$. The orthogonal combination
is given by $U(1)_{Z^\prime}=U(1)_C-U(1)_L$.
In the observable sector there are six horizontal $U(1)$
symmetries.
The first three, $U(1)_{r_j}$ $(j=1,2,3)$, correspond to the right--moving
world--sheet currents ${\bar\eta}_1{\bar\eta}_1^*$,
${\bar\eta}_2{\bar\eta}_2^*$ and ${\bar\eta}_3{\bar\eta}_3^*$.
The last three, $U(1)_{r_{j+3}}$ $(j=1,2,3)$,
correspond to the right--moving
world--sheet currents,
${\bar y}^3{\bar y}^6$, ${\bar y}^1{\bar\omega}^5$ and
${\bar\omega}^2{\bar\omega}^4$ respectively.
For every right--moving $U(1)$ symmetry there is
a left--moving global $U(1)$ symmetry. The first three
correspond to the charges of the supersymmetry generator
$\chi^{12}$, $\chi^{34}$ and $\chi^{56}$. The last three,
$U(1)_{\ell_{j+3}}$ $(j=1,2,3)$,
correspond to the complexified left--moving fermions
$y^3y^6$, $y^1\omega^5$ and $\omega^2\omega^4$.
Finally the model contains six Ising model sigma operators
which are obtained by pairing a left--moving
real fermion with a right--moving real fermion,
$\sigma^i_\pm=\{\omega^1{\bar\omega}^1,
y^2{\bar y}^2, \omega^3{\bar\omega}^3, y^4{\bar y}^4,
y^5{\bar y}^5, \omega^6{\bar\omega}^6\}_\pm$.

The full massless spectrum and cubic level superpotential were presented
in Ref. [\EU]. Here we list only the states which are relevant for the
neutrino see--saw mass matrix.
The sectors $b_1$, $b_2$ and $b_3$ produce
three generations of chiral fermions, $G_1$, $G_2$ and $G_3$, with
horizontal symmetries. For every generation, $G_j$ there are
two right--moving, $U(1)_{r_j}$ and $U(1)_{r_{j+3}}$, symmetries.
For every right--moving $U(1)$ gauged symmetry, there is a corresponding
left--moving global $U(1)$ symmetry, $U(1)_{\ell_j}$ and
$U(1)_{\ell_{j+3}}$. Each sector $b_1$, $b_2$ and $b_3$
has two Ising model operators,
($\sigma_4$, $\sigma_5$), ($\sigma_2$, $\sigma_6$) and ($\sigma_1$,
$\sigma_3$),
respectively, obtained by pairing a left--handed real fermion with
a right--handed real fermion. The pairings of real fermions, which produce
the Ising model operators, is noted in Table 1. The
nonvanishing cubic level and higher order terms in the superpotential
must be invariant under all the symmetries of the superstring model.
The Ising model operators must produce nonvanishing corralators
after all picture changing operations have been performed.

The Neveu--Schwarz sector  produces three pairs of electroweak doublets
$h_1$, ${\bar h}_1$, $h_2$, ${\bar h}_2$ and $h_3$, ${\bar h}_3$. Each pair
carries $(U(1)_{\ell_j};U(1)_{r_j})$ charges, $(j=1,2,3)$, respectively.
The Neveu--Schwarz sector also produces three pairs of $SO(10)$ singlets
$\Phi_{12}$, ${\bar\Phi}_{12}$, $\Phi_{13}$, ${\bar\Phi}_{13}$,
$\Phi_{23}$, ${\bar\Phi}_{23}$ with $U(1)_{r_j}$ charges.
Finally, the Neveu--Schwartz sector gives rise to three singlet
states that are neutral under all the U(1) symmetries.
$\xi_{1,2,3}:{\hskip .2cm}{\chi^{12}_{1\over2}{\bar\omega}^3_{1\over2}
{\bar\omega}^6_{1\over2}{\vert 0\rangle}_0},$
 ${\chi^{34}_{{1\over2}}{\bar y}_{1\over2}^5{\bar\omega}_{1\over2}^1
{\vert 0\rangle}_0},$
 $\chi^{56}_{1\over2}{\bar y}_{1\over2}^2{\bar y}_{1\over2}^4
{\vert 0\rangle}_0.$

The ${S+b_1+b_2+\alpha+\beta}$ sector gives
$$\eqalignno{h_{45}&\equiv{[(1,0);(2,1)]}_
{-{1\over2},-{1\over2},0,0,0,0} {\hskip .5cm}
D_{45}\equiv{[(3,-1);(1,0)]}_
{-{1\over2},-{1\over2},0,0,0,0}&(2a,b)\cr
\Phi_{45}&\equiv{[(1,0);(1,0)]}_
{-{1\over2},-{1\over2},-1,0,0,0}  {\hskip .5cm}
\Phi^{\pm}_1\equiv{[(1,0);(1,0)]}_
{-{1\over2},{1\over2},0,\pm1,0,0}&(2c,d)\cr
\Phi^{\pm}_2&\equiv{[(1,0);(1,0)]}_
{-{1\over2},{1\over2},0,0,\pm1,0} {\hskip .5cm}
\Phi^{\pm}_3\equiv{[(1,0);(1,0)]}_
{-{1\over2},{1\over2},0,0,0,\pm1}&(2e,f)\cr}$$
(and their conjugates ${\bar h}_{45}$, etc.).
The states are obtained by acting on the vacuum
with the fermionic oscillators
${\bar\psi}^{4,5},{\bar\psi}^{1,...,3},{\bar\eta}^3,{\bar y}^3\pm
i{\bar y}^6,{\bar y}^1\pm{i{\bar\omega}^5},
{\bar\omega}^2{\pm}i{\bar\omega}^4$,
respectively  (and their complex conjugates for ${\bar h}_{45}$, etc.).

The sectors $b_i+2\gamma+(I){\hskip .2cm} (i=1,..,3)$ give vector
representations which are $SU(3)_C\times SU(2)_L\times {U(1)_L}\times {U(1)_C}$
singlets (see Table 2). These states transform in the fundamental
representation under the hidden non--Abelian gauge groups.
The sectors $b_1+b_3+\alpha+\gamma+(I)$ and  $b_2+b_3+\beta+\gamma+(I)$
produce states with vanishing weak hypercharge and with non vanishing
$U(1)_{Z^\prime}$ charge (see Table 3). In particular the sector
$b_1+b_3+\alpha+\gamma+(I)$ produces $5$ and ${\bar 5}$ of the hidden
$SU(5)$ gauge group with $U(1)_{Z^\prime}$ charge and vanishing
weak hypercharge. However, the entire massless spectrum does not
contain $5$ and $\bar5$ of the hidden $SU(5)$ group that carry $U(1)_Y$
charge.

The standard--like models contain
an anomalous $U_A(1)$ symmetry which induces a D--term
and destabilizes the vacuum [\DSW]. To preserve supersymmetry
at the Planck scale,
one must satisfy the F and D constraints arising from the
superpotential by giving VEVs to the scalar fields [\ADS].
We will consider the solution to the constraints given by Eq. 12
of Ref. [\EU]:
$$\vert \Phi_{45} \vert^2 =3 \Delta_{13} =3\vert \bar
\Phi_1^- \vert^2 =3\vert \bar \Phi_2^- \vert^2 =3\vert \Phi_3^+ \vert^2=
{g^2 \over16 \pi^2}{1 \over \sqrt{2\alpha^ \prime}} \eqno(3) $$
where $\Delta_{13}=\vert \bar \Phi_{13} \vert^2 - \vert \Phi_{13} \vert^2 $
and $\sqrt{2\alpha^{\prime}}=2\sqrt{8\pi}/{gM_{Pl}}$.
We would like to emphasize that the solution, Eq. (3), preserves F
flatness to all orders of nonrenormalizable terms. For solutions that break
$U(1)_{Z^\prime}$ close to the Planck scale, higher order terms are expected
to violate F flatness [\NRT]. We therefore assume that VEVs which break
$U(1)_{Z^\prime}$ are suppressed relative to the $SO(10)$ singlet VEVs.

\bigskip
\centerline {\bf 3. Neutrino mass matrix}

The cubic level and higher order terms in the superpotential are
derived by applying the rules of Ref. [\KLN]. Order $N$ nonrenormalizable
terms are of the form $c ff h (\Phi/M)^{N-3}$ where f, h and $\Phi$ are
fermions, scalar doublets and singlets respectively. M is a Planck scale mass
to be defined later. The coefficients $c$ can be calculated from string
amplitudes and are $\sim O(1)$. For our choice of boundary condition vectors
given in Table 1, at the cubic level, there are Dirac mass terms only for
the $+{2\over3}$ charged quarks and for the
neutrinos, with $\lambda_u=\lambda_\nu$,
$$(
{u_{L_1}^c}Q_1{\bar h}_1+{N_{L_1}^c}L_1{\bar h}_1+
{u_{L_2}^c}Q_2{\bar h}_2+{N_{L_2}^c}L_2{\bar h}_2+
{u_{L_3}^c}Q_3{\bar h}_3+{N_{L_3}^c}L_3{\bar h}_3). \eqno (4)$$

At the cubic level of the superpotential only two pairs of Higgs
doublets which are combinations of {$h_1$,$h_2$, $h_{45}$} and
{$\bar h_1$,$\bar h_2$,$\bar h_{45}$} remain light. At an intermediate
energy scale one additional pair becomes heavy [\NRT]. For a specific choice
of VEVs, the light doublets will be $\bar h_2$ and $ h_{45}$ [\NRT]. Therefore,
there is a cubic level mass term only for the top quark and for the $\tau$
neutrino (the heaviest generation is the one with index 2), with
$\lambda_t=\lambda_{\nu_\tau}$. The $\mu$ neutrino
obtains a Dirac mass term at order $N=5$.
$$\eqalignno{&N_2L_2 (\bar h_1 \bar \Phi^+_i \bar \Phi^-_i + \bar h_{45} \bar
\Phi_{23} \Phi_{45})  &(5a)\cr
      &N_1L_1 (\bar h_2 \Phi^+_i \Phi^-_i +\bar h_{45} \bar \Phi_{13}
\Phi_{45})
      &(5b)\cr }$$
%      &u_2Q_2({\bar h}_{45}\Phi_{45}{\bar\Phi}_{23}+
%  {\bar h}_1{\bar\Phi}_i^+{\bar\Phi}_i^-)&(28c)\cr
%      &u_1Q_1({\bar h}_{45}\Phi_{45}{\bar\Phi}_{13}+
%  {\bar h}_2{\Phi}_i^+{\Phi}_i^-)&(28d)\cr }$$

Since the VEVs of the scalar fields are generally $\sim O(10^{-1})M$ the Dirac
mass term for the $\mu$ neutrino is suppresed relative to the
$\tau$ neutrino by ${\left(\langle\phi\rangle \over M \right)}^2\sim0.01$.
As shown
in Ref. [\GEN] a Dirac mass term for the electron neutrino can only arise
from higher order terms which contain VEVs that break $U(1)_{Z^\prime}$.
VEVs that break $U(1)_{Z^\prime}$ are expected to be suppressed
relative to the $SO(10)$ singlet VEVs [\NRT].

Mixing between the generations is induced by the states from the sectors
$b_j+2\gamma$ [\CAB].
However, the mixing terms are suppressed relative to the leading diagonal
terms and therefore their effect is small. We neglect them in this paper.

In the superstring derived standard--like model the states in the see--saw
mass matrix are the neutral states in the left--handed doublets $L_1$,
$L_2$, $L_3$, the right--handed neutrinos $N_1$, $N_2$, $N_3$, from the
sectors $b_1$, $b_2$ and $b_3$ and the $SO(10)$ singlets $\Phi=
\{\Phi_{13}, \bar \Phi_{13}, \Phi_1^+, \bar \Phi_1^-, \Phi_2^+, \Phi_3^+,
\bar \Phi_3^-\}$
from the Neveu--Schwarz sector and from the sector $S+b_1+b_2+\alpha+\beta$.
These $SO(10)$ singlets remain massless after the symmetry breakdown for the
choice of VEVs given by Eq. (3) and therefore appear in the low energy
spectrum.
There are other light
singlets but only these mix with the right--handed neutrinos.
A nonvanishing VEV for the right--handed neutrino will induce rapid proton
decay from dimension four operators [\NRT].
Therefore, the right--handed neutrinos are assumed to have vanishing
VEVs.

The neutrino--singlet mixing terms arise from higher order nonrenormalizable
terms, which are obtained according to the rules of Ref. [12]. The general form
of the terms is fixed by the symmetries of the model. The coefficients of these
terms can in principle be calculated from string scattering amplitudes,
and are of order one [\KLN].
The nonrenormalizable terms contributing to the see--saw mass matrix
appear at orders $N=6$, $N=7$ and $N=8$.

At order $N=6$ there is only one see--saw term for $N_2$:
$$N_2 \Phi_2^+ H_{19} \bar T_2 H_{23} \Phi_{45} \eqno(6) $$

At order $N=7$ we obtain the following see--saw terms for $N_3$:
$$\eqalignno{&N_3 \bar\Phi_3^- H_{19} \bar T_3 H_{25} \Phi_{45} \bar \Phi_{13}
&(7a) \cr
&N_3 \Phi_3^+ H_{19} \bar T_3 H_{23}  \Phi_{45} \xi_2   &(7b) \cr}$$

Other see--saw terms that appear at this order are suppressed relative to the
ones above since they contain more VEVs that break $U(1)_{Z^\prime}$.
For $N_1$ the mixing terms are at order $N=8$:
$$\eqalignno{& N_1 \Phi^+_1 H_{19} \bar T_1  H_{23} \Phi_{45} \xi_2 \xi_2 &(8a)
\cr
&N_1 \Phi_{45}  H_{19} \bar T_1  H_{25} \Phi^-_1 \bar \Phi_{13} \xi_2 &(8b) \cr
&N_1 \bar \Phi_{13} H_{19} \bar T_1  H_{25} \bar \Phi^-_1 \Phi_{45} \xi_2 &(8c)
\cr
&N_1 \bar \Phi_{13} H_{19} \bar T_1  H_{25} \bar \Phi^-_1 \Phi_{45} \xi_2 &(8d)
\cr
&N_1 \bar \Phi_1^- H_{24}  \bar T_1  H_{25} \Phi_{45} \bar \Phi_{13} \xi_2 &
(8e) \cr}$$

The see--saw terms that are suppressed relative to those above are
neglected. Up to order $N=8$ there are no mass terms for the $SO(10)$ singlets
for our specific choice of VEVs in Eq. (3). At order $N=9$, there is a mass
term of the
form $\Phi_2^+ \Phi_2^+ \bar T_2 \bar T_2 T_2 T_2 \Phi_{45} \Phi_{45}  \xi_2
M^{-6}$ for $\Phi_2^+$. This term contains four VEVs that break $U(1)_
{Z^\prime}$. Other singlets get masses from higher order ($N>9$) terms.

In Eqs. (6,7,8) the $5, \bar 5$ states, $H_{19}$ and $\bar T$ of the
hidden $SU(5)$ group form condensates. These condensates carry nonzero
$U(1)_{Z^\prime}$ charges and therefore will break $U(1)_{Z^\prime}$. In
addition we make the assumption that additional singlets $(H_{23}, H_{25},
H_{26})$ with nonzero $U(1)_{Z^\prime}$ charge get a VEV at the same scale.
The $SO(10)$ singlets from the Neveu--Shwarz sector and the sector $S+b_1+
b_2+\alpha+\beta$ obtain  VEVs by the cancellation of the anomalous D--term
equation. As a result the terms in Eqs. (5,6,7) induce the effective see--saw
terms of the form
$$ N \Phi \left( \Lambda_{Z^\prime}\over M \right)^3\left( {\langle \phi
\rangle} \over M \right)^n M \eqno(9)$$
where $\Lambda_{Z^\prime}$ is the scale at which $SU(5)$ condensates form and
break $U(1)_{Z^\prime}$ and $M=M_P/2\sqrt {8\pi} \sim 10^{18}$ GeV. $n$
is 1,2 and 3 for $N_2$, $N_3$ and $N_1$ respectively.
We take the $SO(10)$ singlet mass terms, $m_\phi$ to be generically of the form
$$\Phi \Phi \left(\Lambda_{Z^\prime} \over M \right)^4 \left(\langle \phi
\rangle \over M \right)^m M \eqno(10) $$
where $m=3$ for $\Phi_2^+$ and $m>3$ for the other singlets.
It should be emphasized that for the see--saw mechanism to operate there must
exist  singlets that couple to the right--handed neutrinos as in Eq. (9) and
that remain massless at trilevel for the flat directions. If such singlets
exist
then the higher order mass terms for these singlets will be generically of
the form given by Eq. (10).

The neutrino mass matrix therefore takes the following form for each generation
in the basis $(\nu_L, N^C, \Phi)$
$$\left( \matrix{0&km_u&0 \cr
          km_u&0&m_\chi \cr
          0&m_\chi&m_\phi \cr} \right) \eqno(11)$$
with $m_\chi \sim  \left( \Lambda_{Z^\prime}\over M \right)^3\left({\langle
\phi \rangle }\over M \right)^n M$ and $m_\phi \sim \left(\Lambda_{Z^\prime}
\over M \right)^4 \left(\langle \phi \rangle \over M \right)^m M$.
The mass eigenstates are mainly $\nu$, $N$ and $\phi$ with a small mixing and
with the eigenvalues
$$m_{\nu} \sim m_\phi \left({{k m_u} \over m_{\chi}}\right)^2
\qquad m_N,M_{\phi} \sim m_{\chi} \eqno(12)$$

The constant $k$ gives the effects of Yukawa coupling renormalization. At $M$
$\lambda_u=\lambda_{\nu}$. These two couplings run differently since up--quarks
and neutrinos have different quantum numbers. In order to find their ratio at
the weak scale,$M_W$, one has to solve the renormalization group equations for
these two couplings which are
$${d\lambda_1^t \over dt} ={\lambda_1^t \over 32\pi^2} \left (12\lambda_1^{t2}+
2\lambda_2^{b2}+2\lambda_4^2+2\lambda_7^2-4\left ({8\over 3}g_3^2+{3\over 2}
g_2^2+{13\over 30}g_1^2\right)\right) \eqno(13)$$
$${d\lambda_4 \over dt}={\lambda_4 \over 8\pi^2} \left(2\lambda_4^2+{3\over 2}
\lambda_1^{t2}+{1\over 2}\lambda_3^{\tau 2}+{1\over 2}\lambda_7^2-\left (
{3\over 2} g_2^2+{3 \over 10}g_1^2\right)\right) \eqno (14)$$
where $\lambda_1^t,\lambda_2^b,\lambda_3^{\tau},\lambda_4,\lambda_7$ are the
Yukawa couplings of the top, bottom,$\tau$, $\tau$ neutrino and scalar Higgs
respectively. $g_1$, $g_2$ and $g_3$ are the $U(1)_Y$, $SU(2)_L$ and $SU(3)_C$
gauge couplings. The numerical solution gives $k \sim 0.5$.

The ratio $(\Lambda_{Z^\prime}/ M)$ is obtained from the one loop
renormalization group equations
$$\left( \Lambda_{Z^\prime}\over M \right)=exp\left(32\pi(1-\alpha_5(M))
\over b_5 \right) \eqno(15)$$
with $b_5={1\over 2}n_5-15$,  $\alpha_5(M) \sim 0.059$ and
$n_5=8$ which give $\Lambda_{Z^\prime} \sim 10^{14} GeV$.
Taking the $SO(10)$ singlet VEVs to be $\sim 10^{-1}M$, a numerical scenario
for the heavy generation gives
$$m_{\nu_\tau} \sim 25~eV  \qquad  m_{N_\tau} \sim 100~TeV \eqno(16)$$
Assuming that other singlet masses arise at order $N=10$ (or $m=4$), one gets
the loose bounds for the other two generations:
$$m_{\nu_e}, m_{\nu_\mu} < O(10^{-4}~ eV)\quad
m_{N_e}, m_{N_\mu} > O(1~ TeV) \eqno(17)$$

If some of the $5,\bar 5$ states of the hidden $SU(5)_H$ get mass after the
symmetry breakdown, $n_5$
gets smaller. For smaller $n_5$, $\Lambda_{Z^\prime}$ and the see-saw scale
$m_\chi$ increase. Therefore the left-handed neutrino masses decrease and the
right--handed neutrino masses increase. It is also possible to
modify the boundary conditions of the hidden sector in the vector $\gamma$:
$\gamma(\phi^{1,2 \cdots 8})=({1\over 2} 0 0 0 {1\over 2} {1\over 2}
{1\over 2} 0)$ [\CAB].
Then the hidden gauge group is $SU(7) \times U(1)^2$ in which case
$\Lambda_{Z^\prime}$ is larger and the left--handed neutrino masses are
smaller.

\bigskip
\centerline{\bf 4.  Conclusions }

We have shown that although there is a large Dirac mass for the $\tau$--
neutrino at tree level in the superpotential of the superstring
standard--like models, very
light left--handed and heavy right--handed neutrinos for all generations are
obtained.
This is due to the see--saw mechanism present in the model as a consequence
of the higher order nonrenormalizable terms
and the breaking of $U(1)_{Z^\prime}$ symmetry by hidden sector
condensates. The hidden sector condensates break $U(1)_{Z^\prime}$
and leave $U(1)_Y$ unbroken.
We learn from our superstring
derived see--saw mechanism a
new feature unprecedented in traditional GUT models. Although the scale of
$U(1)_{Z^\prime}$ breaking can be large (e.g. $\Lambda_{Z^\prime} \approx 10^
{14} GeV$) the effective see-saw scale is much smaller. This feature arises
in superstring models because the see--saw and the singlet mass terms are
obtained from higher order nonrenormalizable terms. As a result, the
right--handed neutrino masses are suppressed relative to the $U(1)_{Z^\prime}$
breaking scale. Consequently, the right--handed neutrinos can be at a rather
low energy scale, say a few TeV, but this does not imply the existence of a
new gauge boson $(Z^\prime)$ at the TeV scale [\FN].

%\bigskip
\vfill
\eject
\centerline{\bf Acknowledgments}

This work is supported in part by a Feinberg School Fellowship.
Alon Faraggi would like to thank the SSC laboratory for its hospitality
while part of this work was conducted. Edi Halyo
would like to express his gratitude to Isak and Gulen Halyo
for their support during this work.

\refout

\vfill
\eject

\input tables.tex

%\special{landscape}
%\hoffset=1.25truein
%\nopagenumbers
%\magnification=1000
%\font\normalroman=cmr10
%\font\style=cmr7
%\style
\tolerance=1200

%\fontdimen12\fivesy=0pt

%\textfont0=\sevenrm
%\scriptfont0=\fiverm
%\textfont1=\seveni
%\scriptfont1=\fivei
%\textfont2=\sevensy
%\scriptfont2=\fivesy

{\hfill
{\begintable
\  \ \|\ ${\psi^\mu}$ \ \|\ $\{{\chi^{12};\chi^{34};\chi^{56}}\}$\ \|\
{${\bar\psi}^1$, ${\bar\psi}^2$, ${\bar\psi}^3$,
${\bar\psi}^4$, ${\bar\psi}^5$, ${\bar\eta}^1$,
${\bar\eta}^2$, ${\bar\eta}^3$} \ \|\
{${\bar\phi}^1$, ${\bar\phi}^2$, ${\bar\phi}^3$, ${\bar\phi}^4$,
${\bar\phi}^5$, ${\bar\phi}^6$, ${\bar\phi}^7$, ${\bar\phi}^8$} \crthick
$\alpha$
\|\ 0 \|
$\{0,~0,~0\}$ \|
1, ~~1, ~~1, ~~0, ~~0, ~~0 ,~~0, ~~0 \|
1, ~~1, ~~1, ~~1, ~~0, ~~0, ~~0, ~~0 \nr
$\beta$
\|\ 0 \| $\{0,~0,~0\}$ \|
1, ~~1, ~~1, ~~0, ~~0, ~~0, ~~0, ~~0 \|
1, ~~1, ~~1, ~~1, ~~0, ~~0, ~~0, ~~0 \nr
$\gamma$
\|\ 0 \|
$\{0,~0,~0\}$ \|
{}~~$1\over2$, ~~$1\over2$, ~~$1\over2$, ~~$1\over2$,
{}~~$1\over2$, ~~$1\over2$, ~~$1\over2$, ~~$1\over2$ \| $1\over2$, ~~0, ~~1,
{}~~1,
{}~~$1\over2$,
{}~~$1\over2$, ~~$1\over2$, ~~0 \endtable}
\hfill}
\smallskip
{\hfill
{\begintable
\  \ \|\
${y^3y^6}$,  ${y^4{\bar y}^4}$, ${y^5{\bar y}^5}$,
${{\bar y}^3{\bar y}^6}$
\ \|\ ${y^1\omega^6}$,  ${y^2{\bar y}^2}$,
${\omega^5{\bar\omega}^5}$,
${{\bar y}^1{\bar\omega}^6}$
\ \|\ ${\omega^1{\omega}^3}$,  ${\omega^2{\bar\omega}^2}$,
${\omega^4{\bar\omega}^4}$,  ${{\bar\omega}^1{\bar\omega}^3}$  \crthick
$\alpha$ \|
1, ~~~0, ~~~~0, ~~~~0 \|
0, ~~~0, ~~~~1, ~~~~1 \|
0, ~~~0, ~~~~1, ~~~~1 \nr
$\beta$ \|
0, ~~~0, ~~~~1, ~~~~1 \|
1, ~~~0, ~~~~0, ~~~~0 \|
0, ~~~1, ~~~~0, ~~~~1 \nr
$\gamma$ \|
0, ~~~1, ~~~~0, ~~~~1 \|\
0, ~~~1, ~~~~0, ~~~~1 \|
1, ~~~0, ~~~~0, ~~~~0  \endtable}
\hfill}
\smallskip
\parindent=0pt
\hangindent=39pt\hangafter=1
%\normalroman
\baselineskip=18pt

{{\it Table 1.} A three generations ${SU(3)\times SU(2)\times U(1)^2}$
model. The choice of generalized GSO coefficients is:
${c\left(\matrix{b_j\cr
                                    \alpha,\beta,\gamma\cr}\right)=
-c\left(\matrix{\alpha\cr
                                    1\cr}\right)=
c\left(\matrix{\alpha\cr
                                    \beta\cr}\right)=
-c\left(\matrix{\beta\cr
                                    1\cr}\right)=
c\left(\matrix{\gamma\cr
                                    1,\alpha\cr}\right)=
-c\left(\matrix{\gamma\cr
                                    \beta\cr}\right)=
-1}$ (j=1,2,3),
with the others specified by modular invariance and space--time
supersymmetry.
Trilevel Yukawa couplings are obtained only for
${+{2\over3}}$ charged quarks.
The $16$ right--moving
internal fermionic states
$\{{\bar\psi}^{1,\cdots,5},{\bar\eta}^1,
{\bar\eta}^2,{\bar\eta}^3,{\bar\phi}^{1,\cdots,8}\}$,
correspond to the $16$ dimensional compactified  torus of the ten dimensional
heterotic string.
The 12 left--moving and 12 right--moving real internal fermionic states
correspond to the six left
and six right compactified dimensions in the bosonic language.
$\psi^\mu$ are the two space--time
external fermions in the light--cone gauge and
$\chi^{12}$, $\chi^{34}$, $\chi^{56}$
correspond to the spin connection in the bosonic constructions.}
\vskip 2.5cm

\vfill
\eject

\baselineskip=18pt
\hbox
{\hfill
{\begintable
\ F \ \|\ SEC \ \|\ $SU(3)_C$ $\times$ $SU(2)_L$ \ \|\ $Q_C$ & $Q_L$ & $Q_1$ &
$Q_2$
 & $Q_3$ & $Q_4$  & $Q_5$ & $Q_6$
 \ \|\ $SU(5)$ $\times$ $SU(3)$ \ \|\ $Q_7$ &
$Q_8$  \crthick
$V_1$ \|\ ${b_1+2\gamma}+(I)$ \|(1,1)\|~0 & ~~0 & ~~0 & ~~${1\over 2}$ &
 ~~$1\over 2$ & ~~$1\over2$
 & ~~0 & ~~0 \|(1,3)\| $-{1\over 2}$ &
{}~~$5\over 2$   \nr
${\bar V}_1$ \|\                \|(1,1)\| ~~0 & ~~0 & ~~0 & ~~${1\over 2}$ &
 ~~$1\over 2$ & ~~$1\over2$
  & ~~0 & ~~0
  \|(1,$\bar 3$)\| ~~${1\over 2}$ &
$-{5\over 2}$  \nr
$T_1$ \|\                \|(1,1)\| ~~0 & ~~0 & ~~0 & ~~${1\over 2}$ &
 ~~$1\over 2$ & $-{1\over2}$
  & ~~0 & ~~0
  \|(5,1)\| $-{1\over 2}$ &
$-{3\over 2}$  \nr
${\bar T}_1$ \|\                \|(1,1)\| ~~0 & ~~0 & ~~0 & ~~${1\over 2}$ &
 ~~$1\over 2$ & $-{1\over2}$
   & ~~0 & ~~0
  \|($\bar 5$,1)\| ~~$1\over2$ &
{}~~${3\over 2}$  \cr
$V_{2}$ \|\ ${b_2+2\gamma}+(I)$ \|(1,1)\| ~~0 & ~~0 &
 ~~${1\over 2}$ & ~~0 &
 ~~$1\over 2$ & ~~0 &  ~~$1\over2$ & ~~0
 \|(1,3)\| $-{1\over 2}$ &
 ~~$5\over 2$  \nr
${\bar V}_{2}$ \|\                \|(1,1)\| ~~0 & ~~0 & ~~${1\over 2}$ & ~~0 &
 ~~$1\over 2$ & ~~0 & ~~$1\over2$ & ~~0
  \|(1,$\bar 3$)\| ~~${1\over 2}$ &
 $-{5\over 2}$  \nr
$T_{2}$ \|\                \|(1,1)\| ~~0 & ~~0 & ~~${1\over 2}$ & ~~0 &
 ~~$1\over 2$ & ~~0 & $-{1\over2}$ & ~~0
  \|(5,1)\| $-{1\over 2}$ &
 $-{3\over 2}$ \nr
${\bar T}_{2}$ \|\                \|(1,1)\| ~~0 & ~~0 & ~~${1\over 2}$ & ~~0 &
 ~~$1\over 2$ & ~~0 & $-{1\over2}$ & ~~0
  \|($\bar 5$,1)\| ~~$1\over 2$ &
 ~~${3\over 2}$  \cr
$V_{3}$ \|\ ${b_3+2\gamma}+(I)$ \|(1,1)\| ~~0 & ~~0 & ~~${1\over 2}$ &
 ~~${1\over 2}$ & ~~0 & ~~0 & ~~0 & ~~${1\over2}$
       \|(1,3)\| $-{1\over 2}$ &
 ~~$5\over 2$  \nr
${\bar V}_{3}$ \|\                \|(1,1)\| ~~0 & ~~0 & ~~${1\over 2}$ &
 ~~${1\over 2}$ & ~~0 & ~~0 & ~~0 & ~~$1\over2$
 \|(1,$\bar 3$)\| ~~${1\over 2}$
 & $-{5\over 2}$  \nr
$T_{3}$ \|\                \|(1,1)\| ~~0 & ~~0 & ~~${1\over 2}$ &
 ~~${1\over 2}$ & ~~0 & ~~0 & ~~0 & $-{1\over2}$
       \|(5,1)\|
 $-{1\over 2}$ & $-{3\over 2}$  \nr
${\bar T}_{3}$ \|\                \|(1,1)\| ~~0 & ~~0 & ~~${1\over2}$ &
 ~~${1\over 2}$ & ~~0 & ~~0 & ~~0 & $-{1\over2}$
   \|($\bar 5$,1)\| ~~$1\over 2$
 & ~~${3\over 2}$
 \endtable}
\hfill}
\bigskip
\parindent=0pt
\hangindent=39pt\hangafter=1
{\it Table 1.} Massless states from the sectors $b_j+2\gamma$,
and their quantum numbers.

\vfill
\eject

\baselineskip=18pt
\hbox
{\hfill
{\begintable
\ F \ \|\ SEC \ \|\ $SU(3)_C$ $\times$ $SU(2)_L$ \ \|\ $Q_C$ & $Q_L$ & $Q_1$ &
$Q_2$
 & $Q_3$ & $Q_4$  & $Q_5$ & $Q_6$
 \ \|\ $SU(5)$ $\times$ $SU(3)$ \ \|\ $Q_7$ &
$Q_8$  \crthick
$H_{13}$ \|\ $b_1+b_3+\alpha$  \|(1,1)\| $-{3\over 4}$ &
{}~~$1\over2$
& $-{1\over4}$ & ~~${1\over 4}$ & $-{1\over 4}$ & ~~0
 & ~~0 & ~~0 \|(1,3)\| ~~${3\over 4}$ & ~~$5\over 4$   \nr
$H_{14}$ \|\ $\pm \gamma+(I)$  \|(1,1)\| ~~$3\over4$ & $-{1\over2}$ & ~
$1\over4$
& $-{1\over 4}$ & ~~$1\over 4$ & ~~0 & ~~0 & ~~0
  \|(1,$\bar 3$)\| $-{3\over 4}$ & $-{5\over 4}$  \nr
$H_{15}$ \|\                \|(1,2)\| $-{3\over4}$ & $-{1\over2}$ &
$-{1\over4}$
& ~~${1\over 4}$ & $-{1\over 4}$ & ~~0 & ~~0 & ~~0
  \|(1,1)\| $-{1\over 4}$ & $-{15\over 4}$  \nr
$H_{16}$ \|\                \|(1,2)\| ~~$3\over4$ & ~~$1\over2$ & ~~$1\over4$
& $-{1\over 4}$ & ~~$1\over 4$ & ~~0 & ~~0 & ~~0
  \|(1,1)\| ~~$1\over4$ & ~~${15\over 4}$  \nr
$H_{17}$ \|\                  \|(1,1)\| $-{3\over4}$ & ~~$1\over2$ &
 $-{1\over 4}$ & $-{3\over4}$ & $-{1\over 4}$ & ~~0 & ~0& ~~0
 \|(1,1)\| $-{1\over 4}$ & $-{15\over 4}$  \nr
$H_{18}$ \|\                \|(1,1)\| ~~$3\over4$ & $-{1\over2}$ & ~
${1\over 4}$
& ~~$3\over4$ & ~~$1\over 4$ & ~~0 & ~~0 & ~~0
  \|(1,1)\| ~~${1\over 4}$ & ~~${15\over 4}$  \cr
$H_{19}$ \|\ $b_2+b_3+\alpha$  \|(1,1)\| $-{3\over4}$
& ~~$1\over2$
& ~~${1\over 4}$ & $-{1\over4}$ & $-{1\over 4}$ & ~~0 & ~0& ~~0
  \|(5,1)\| $-{1\over 4}$ & ~~${9\over 4}$ \nr
$H_{20}$ \|\ $\pm \gamma+(I) $    \|(1,1)\| ~~$3\over4$ & $-{1\over2}$ &
$-{1\over 4}$
& ~~$1\over4$ & ~~$1\over 4$ & ~~0 & ~0& ~~0 \|($\bar 5$,1)\| ~~$1\over 4$ &
 $-{9\over 4}$  \nr
$H_{21}$ \|\                 \|(3,1)\| ~~$1\over4$ & ~~$1\over2$ & ~
${1\over 4}$
& $-{1\over 4}$ & $-{1\over4}$ & ~~0 & ~~0 & ~~0
       \|(1,1)\| $-{1\over 4}$ &
 $-{15\over 4}$  \nr
$H_{22}$ \|\                \|($\bar 3$,1)\| $-{1\over4}$ & $-{1\over2}$
& $-{1\over 4}$ & ~~${1\over 4}$ & ~~$1\over4$ & ~~0 & ~~0 & ~~0
 \|(1,1)\| ~~${1\over 4}$ & ~~${15\over 4}$  \nr
$H_{23}$ \|\                \|(1,1)\| $-{3\over4}$ & ~~$1\over2$ & ~
${1\over 4}$
& $-{1\over 4}$ & ~~$3\over4$ & ~~0 & ~~0 & ~~0       \|(1,1)\|
 ~~${1\over 4}$ & ~~${15\over 4}$  \nr
$H_{24}$\|\                \|(1,1)\| ~~$3\over4$ & $-{1\over2}$ & $-{1\over4}$
& ~~${1\over 4}$ & $-{3\over4}$ & ~~0 & ~~0 & ~~0 \|(1,1)\| ~~$1\over 4$
 & ~~${15\over 4}$ \nr
$H_{25}$ \|\                \|(1,1)\| $-{3\over4}$ & ~~$ 1\over2$ &~~$1\over4$
&
{}~~$3\over4$ & $-{1\over4}$ & ~~0 & ~~0 & ~~0 \|(1,1)\| $-{1\over4}$ &
$-{15\over4}$ \nr
$H_{26}$ \|\                \|(1,1)\| ~~$3\over4$ &  $-{1\over2}$ &
$-{1\over4}$ &  $-{3\over4}$ & ~~$1\over4$ & ~~0 & ~~0 & ~~0 \|(1,1)\|
{}~~$1\over4$ & ~~$15\over4$
\endtable}
\hfill}
\bigskip
\parindent=0pt
\hangindent=39pt\hangafter=1
{\it Table 3.} Massless scalars and their quantum numbers

\vfill
\eject

\end
\bye